# Demonstration of Jarzynski's Equality in Open Quantum Systems Using a Step-Wise Pulling Protocol

## Van A. Ngo and Stephan Haas

*Department of Physics and Astronomy, University of Southern California, Los Angeles, California 90089, USA*

We present a generalization of Jarzynski's Equality, applicable to quantum systems, relating discretized mechanical work and free-energy changes. The theory is based on a step-wise pulling protocol. We find that work distribution functions can be constructed from fluctuations of a reaction coordinate along a reaction pathway in the step-wise pulling protocol. We also propose two sets of equations to determine the two possible optimal pathways that provide the most significant contributions to free-energy changes. We find that the transitions along these most optimal pathways, satisfying both sets of equations, follow the principle of *detailed balance*. We then test the theory by explicitly computing the free-energy changes for a one-dimensional quantum harmonic oscillator. This approach suggests a feasible way of measuring the fluctuations to experimentally test Jarzynski's Equality in many-body systems, such as Bose-Einstein condensates.

## I. INTRODUCTION

Jarzynski's Equality (JE) is well-known for closed systems, both in classical and quantum mechanics [1–5]. The JE describes a relation between applied work $W$ and free-energy changes [see Eq. (1)]. For closed quantum systems it does not require any heat bath to maintain temperature. As a result, work $W$ is evaluated as $E_m(t_f) - E_n(0)$, where $E_m(t_f)$ is the $m$-th eigenvalue of a final Hamiltonian $\hat{H}(t_f)$ at time $t_f$, and $E_n(0)$ is the $n$-th eigenvalue of an initial Hamiltonian $\hat{H}(t=0)$ [3]. In contrast, for open quantum systems $W$ cannot be expressed in terms of simple energy differences, due to the presence of heat baths. To extend the JE for open systems, Crooks [6] developed a theory that considers the relation between discretized mechanical work $W = \sum_{i=1}^{s-1}[H(x_i,\lambda_{i+1}) - H(x_i,\lambda_i)]$ and free-energy changes $\Delta F(\lambda_1,\lambda_s)$ for open stochastic classical systems, whose evolution follows the principle of *detailed balance*. Here the systems are characterized by Hamiltonian $H(x,\lambda)$; $x_i$ denotes a reaction coordinate $x$ at the $i^{th}$-discretized step; $\lambda$ is a control parameter; and $s$ is the number of discretized steps. The relation is

$$\exp[-\Delta F(\lambda_1,\lambda_s)/k_{\mathbf{B}}T] = \langle\exp(-W/k_{\mathbf{B}}T)\rangle, \quad (1)$$

where $k_{\mathbf{B}}$ is Boltzmann's constant, $T$ is temperature, and $\langle\ldots\rangle$ indicates the average over all possible values of $W$.

To further examine the connection between work distributions and free-energy changes for general open classical systems, Crooks formulated a fluctuation theorem, which relates the distributions of work $\rho_F(+W)$ in forward processes with $\rho_R(-W)$ in reverse processes via $\rho_F(+W)/\rho_R(-W) = \exp(\beta[W-\Delta F(\lambda_1,\lambda_s)])$ [7]. This theorem is more universal than the JE, because one can obtain the JE by multiplying both sides with $\rho_R(-W)\exp(-\beta W)$ and integrating over $W$. Later, Campisi *et al.* [8] extended the applicability of the JE and the fluctuation theorem for open and arbitrarily strong-coupling quantum systems. However, the expression of work $W = \sum_{i=1}^{s-1}[H(x_i,\lambda_{i+1}) - H(x_i,\lambda_i)]$ has not yet been shown to be essential for open quantum systems. In other words, the possibility of constructing work distribution functions by using the explicit discretized form of $W$ in quantum mechanics has not been studied. Since the major issue of the JE is how to perform and measure work distribution functions [9], this would shed new light on how the JE works in quantum systems.

Recently, one of us [10] presented a proof for the relation between the discretized mechanical work $W = \sum_{i=1}^{s-1}[U(x_i,\lambda_{i+1}) - U(x_i,\lambda_i)]$ and free-energy changes in open classical systems without using the principle of *detailed balance*. This proof is based on a step-wise pulling protocol (see Fig. 1), in which an applied potential $U(x,\lambda)$ is used to perform work $W$ in a step-wise manner. To implement the step-wise pulling protocol, a double Heaviside functions of time $t$, $\theta(t-t_{i-1})\theta(t_i-t)$, was used, where the index $i$ denotes a pulling step. For a relaxation time $t_{i-1} - t_i$ the applied potential is $U(x_i,\lambda=\lambda_i)$, in which $x_i$ varies during the relaxation time. By introducing the double Heaviside functions into a coupled Hamiltonian $H(t)$, one can verify that $\langle\exp(-\int[\partial H(t)/\partial t]dt/k_{\mathbf{B}}T)\rangle = 1$, which helps to prove the JE, Eq. (1). If $H(t)$ does not contain double Heaviside functions of time $t$, one can instead consider Hummer-Szabo's proof [11–14] for the JE. Although these two proofs are different, Ngo, Hummer, and Szabo arrived at similar approaches for reconstructing free-energy landscapes, which take into account the initial positions of an applied potential.

The main advantage of using a step-wise pulling protocol is that it allows to obtain work distribution functions by measuring thermal-fluctuation distributions of $x_i$. These distributions of $x_i$ can be generated with a finite number $s$ of pulling steps, and reasonably small relaxation times $(t_i - t_{i-1})$, hence utilizing non-equilibrium pulling processes. Finite relaxation times are required to allow the system to evolve into states, which are used to generate distributions of trajectories along a reaction pathway. Here, rare trajectories corresponding to small values of work are found to yield the dominant contributions in Eq. (1), when computing free-energy changes. They are present in Eq. (1) if overlaps between successive distributions of fluctuating $x_i$ are larger than the standard deviations of their distribution functions.

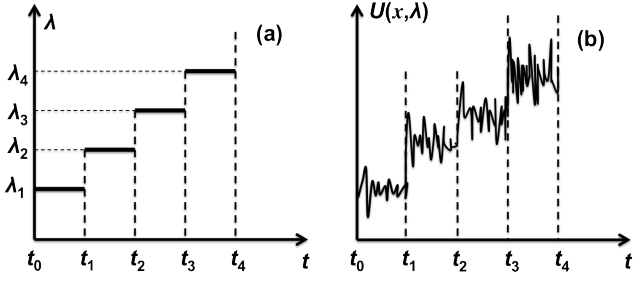

**Fig. 1.** Schematic diagram for a step-wise pulling protocol: (a) control parameter $\lambda$, and (b) expectation value of the applied potential versus time.

In this article, we extend Ngo's proof [9] to quantum systems (Sec. II). Specifically, we define a reference free energy for the free-energy changes and obtain the average on the right-hand side of Eq. (1) in terms of operators. This quantum version of the JE can be applied to any quantum systems, as long as an external potential with a control parameter $\lambda$ is applicable. As an example, using a harmonic potential to perform work, we show how discretized mechanical work enters Eq. (1) for quantum systems. As a result, work distribution functions can be generated from eigenstates and eigenvalues of a coupled Hamiltonian operator, or in general thermal and quantum fluctuation distributions of a reaction coordinate. From an explicit expression of the work distribution function, we obtain two sets of equations to determine which transition pathways provide the dominant contributions to free-energy changes. The pathways satisfying both sets of equations follow the principle of *detailed balance*. We test this theory on a quantum harmonic oscillator (Sec. III). Finally, we discuss some important consequences of this theory (Sec. IV).

## II. THEORY

Let us consider a stationary system of $N$ particles described by a time-independent Hamiltonian $\hat{H}_0(\hat{p}^{3N}, \hat{r}^{3N-1}, \hat{x})$, where $\hat{p}^{3N}$ and $\hat{r}^{3N-1}, \hat{x}$ are $3N$ momentum and position operators, and the system is coupled to a heat bath at temperature $T$. Here $\hat{x}$ is a one-dimensional reaction coordinate operator, which can be coupled to an external potential operator $\hat{U}(\hat{x}, \lambda)$, where the parameter $\lambda$ controls the center of the applied potential. We define a pre-defined classical pathway in which $\lambda$ is changed from $\lambda_1 \rightarrow \lambda_2 \rightarrow \lambda_3 \ldots \rightarrow \lambda_s$, where $s \geq 2$ is the number of pulling steps (see Fig. 1). For each pull, parameterized by $\lambda_i$, the coupled system is allowed to relax. The relaxation time $\tau_i = t_i - t_{i-1}$ is chosen sufficiently large to equilibrate the system following each instantaneous pulling step. This is called a step-wise pulling protocol.

To describe the coupling between $\hat{U}(\hat{x}, \lambda)$ and $\hat{H}_0(\hat{p}^{3N}, \hat{r}^{3N-1}, \hat{x})$, we use double Heaviside functions in time $\theta(t - t_{i-1})\theta(t_i - t)$. Then, at the $i$th-pulling step the coupled Hamiltonian can be written as $\hat{H}(\hat{p}^{3N}, \hat{r}^{3N-1}, \hat{x}; \lambda_i)$ $= \hat{H}_0(\hat{p}^{3N}, \hat{r}^{3N-1}, \hat{x}) + \hat{U}(\hat{x}, \lambda_i)\theta(t - t_{i-1})\theta(t_i - t)$. For sufficiently large $\tau_i$, $\hat{H}(\hat{p}^{3N}, \hat{r}^{3N-1}, \hat{x}; \lambda_i)$ acts as a quasi-time-independent operator. Therefore, we can assume that a canonical ensemble with Hamiltonian $\hat{H}(\hat{p}^{3N}, \hat{r}^{3N-1}, \hat{x}; \lambda_i)$ exists for each pulling step. The general coupled Hamiltonian can be written as

$$\hat{H}(\hat{p}^{3N}, \hat{r}^{3N-1}, \hat{x}; \lambda_1, ..., \lambda_s) = \hat{H}_0(\hat{p}^{3N}, \hat{r}^{3N-1}, \hat{x})$$
$$+ \sum_{i=1}^{s} \hat{U}(\hat{x}, \lambda_i)\theta(t - t_{i-1})\theta(t_i - t). \quad (2)$$

Next, we show how to extract mechanical work from the general coupled Hamiltonian. Let us formally define an operator $\hat{O}_{\text{total}}$ and the mechanical work $\hat{W}$ operator in the step-wise protocol as

$$\hat{O}_{\text{total}} = \int_{t_0}^{t_s} \frac{\partial \hat{H}(\hat{p}^{3N}, \hat{r}^{3N-1}, \hat{x}; \lambda_1, \lambda_1 ... \lambda_s)}{\partial t} dt, \quad (3a)$$

$$\hat{W} = \sum_{i=1}^{s-1} \delta \hat{W}_i = \sum_{i=1}^{s-1} [\hat{U}(\hat{x}, \lambda_{i+1}) - \hat{U}(\hat{x}, \lambda_i)], \quad (3b)$$

where $\delta \hat{W}_i = \hat{U}(\hat{x}, \lambda_{i+1}) - \hat{U}(\hat{x}, \lambda_i)$. (See Appendix A for the relation between $\hat{O}_{\text{total}}$ and $\hat{W}$). Let us now denote $x_i$ to be the eigenvalues of the operator $\hat{x}$ for the $i$th-pulling step from $t_{i-1}$ to $t_i$. Then, for a trajectory $x_1 \rightarrow x_2 \rightarrow x_3 \ldots \rightarrow x_{s-1}$ we can write the expression of the expectation value $W$ (without hat) as follows:

$$W = \sum_{i=1}^{s-1} [U(x_i, \lambda_{i+1}) - U(x_i, \lambda_i)]. \quad (3c)$$

We will now show how this expression for $W$ can be used to generate a work distribution function using $s - 1$ distributions of fluctuating $x_i$ to evaluate the average in the Jarzynski's Equality Eq. (1) for quantum systems.

To derive Jarzynski's Equality Eq. (1), we have to specify how to take averages for pre-defined classical pathways, and define a reference free energy. Firstly, we define $C(\lambda_1, \ldots, \lambda_s)$ as a function of only $(\lambda_1, \ldots, \lambda_s)$:

$$C(\lambda_1, ..., \lambda_s) = \{Z_0 \prod_{i=1}^{s} Z_i\}^{-1} \times \text{Tr}_{X_0} e^{-\beta \hat{H}_0(\hat{p}^{3N}, \hat{r}^{3N-1}, \hat{x})} e^{-\beta \hat{U}(\hat{x}, \lambda_1)}$$
$$\times \prod_{i=1}^{s-1} \left\{ \text{Tr}_{X_i} e^{-\beta[\hat{H}_0(\hat{p}^{3N}, \hat{r}^{3N-1}, \hat{x}) + \hat{U}(\hat{x}, \lambda_i)]} e^{-\beta \delta \hat{W}_i} \right\}$$
$$\times \text{Tr}_{X_s} e^{-\beta \hat{H}_0(\hat{p}^{3N}, \hat{r}^{3N-1}, \hat{x})} e^{+\beta \hat{U}(\hat{x}, \lambda_s)}, \quad (4)$$

where here $\beta$ is $1/k_\text{B}T$; Tr denotes a trace over a complete set of states in the Hilbert space; $X_0$ represents the complete set of states for the initial Hamiltonian $\hat{H}_0(\hat{p}^{3N}, \hat{r}^{3N-1}, \hat{x})$; $X_i$ represents the complete set of states at the $i$th-pulling step for Hamiltonian $\hat{H}_i(\hat{p}^{3N}, \hat{r}^{3N-1}, \hat{x}; \lambda_i) = \hat{H}_0(\hat{p}^{3N}, \hat{r}^{3N-1}, \hat{x})$ $+\hat{U}(\hat{x}, \lambda_i)$. The partition function before pulling is $Z_0 =$ $\text{Tr}_{X_0} \exp[-\beta \hat{H}_0(\hat{p}^{3N}, \hat{r}^{3N-1}, \hat{x})] = \exp[-\beta F_0]$, where $F_0$ is the free energy without the applied potential. The partition function $Z_i$ is $\text{Tr}_{X_i} \exp[-\beta \hat{H}_i(\hat{p}^{3N}, \hat{r}^{3N-1}, \hat{x}; \lambda_i)] = \exp[-\beta F(\lambda_i)]$, where $F(\lambda_i)$ is the free energy at the $i$th-pulling step. Henceforth, we omit $\hat{p}^{3N}, \hat{r}^{3N-1}$ in the Hamiltonian operators to simplify the notation.

We then define the reference free energy $F_{\text{ref}}(\lambda_1, \lambda_s)$ as

$$F_{\text{ref}}(\lambda_1, \lambda_s) = -\beta^{-1} \ln\left( \frac{\text{Tr}_{X_0} e^{-\beta \hat{H}_0(\hat{x})} e^{-\beta \hat{U}(\hat{x}, \lambda_1)}}{C(\lambda_1, ..., \lambda_s)} \right. \tag{5}$$

$$\left. \times \frac{\text{Tr}_{X_s} e^{-\beta [\hat{H}_0(\hat{x}) + \hat{U}(\hat{x}, \lambda_s)]} e^{+\beta \hat{U}(\hat{x}, \lambda_s)}}{Z_0} \right).$$

Substituting the traces over $X_0$ and $X_s$ in Eq. (4) by using Eq. (5), we obtain

$$\exp[-\beta \Delta F(\lambda_1, \lambda_s)] = \exp\left(-\beta [F(\lambda_s) - F_{\text{ref}}(\lambda_1, \lambda_s)]\right)$$
$$= \prod_{i=1}^{s-1} \text{Tr}_{X_i} \hat{\Omega}_i \exp(-\beta \delta \hat{W}_i), \tag{6}$$

which is a quantum-mechanical analogue of the JE. Here, $\hat{\Omega}_i = \exp[-\beta \hat{H}_i(\hat{p}^{3N}, \hat{r}^{3N-1}, \hat{x}; \lambda_i)] / Z_i$ are density operators. In Eq. (6) $F_{\text{ref}}(\lambda_1, \lambda_s)$ depends on $\lambda_s$. In other words, $F_{\text{ref}}(\lambda_1, \lambda_s)$ varies with respect to $s$. Note that in classical mechanics $C(\lambda_1, ..., \lambda_s)$ is identical to unity [10], and $F_{\text{ref}}(\lambda_1, \lambda_s)$ is identical to $F(\lambda_1)$ since the initial Hamiltonian commutes with the applied potential. In quantum mechanics, we will explicitly examine below how $\Delta F(\lambda_1, \lambda_s)$ and $F_{\text{ref}}(\lambda_1, \lambda_s)$ vary in a one-dimensional harmonic oscillator as one keeps $\lambda_s - \lambda_1$ unchanged and increases $s$ at various temperatures (see Sec. III).

To illustrate that Eq. (6) is the operator expression for the JE Eq. (1), we consider a harmonic potential $\hat{U}(\hat{x}, \lambda) = k(x - \lambda)^2 / 2$ to perform work. Then, the expectation value of the mechanical work $W$ following Eq. (3c) is simply a linear function of all $x_i$, which denotes the eigenvalues of $\hat{x}$ at the $i$th-pulling steps. Suppose that at the $i$th-pulling step there is a complete set of states $\{|E_i\rangle\}$ in Hilbert space, which are eigenstates of Hamiltonian $\hat{H}_i(\hat{x}; \lambda_i)$ with eigenvalues $E_i$. In the position representation, each state $|E_i\rangle$ is related to a wave function $\langle \vec{r}^{3N-1}, x_i | E_i \rangle$. Then, the probability of finding the reaction coordinate at $x_i$ in the domain of spatial fluctuations $V_i$ during the $i$th-pulling step is given by $|\psi_{E_i}(x_i)|^2 = \int d\vec{r}^{3N-1} |\langle \vec{r}^{3N-1}, x_i | E_i \rangle|^2$. By using $s - 1$ sets of the states and probabilities $|\psi_{E_i}(x_i)|^2$, and the traces in Eq. (6) can be written as

$$\sum_{E_1, E_2...E_{s-1}} \int_{V_1, V_2...V_{s-1}} \frac{\prod_{i=1}^{s-1} dx_i \prod_{i=1}^{s-1} |\psi_{E_i}(x_i)|^2 e^{-\beta \sum_{i=1}^{s-1} E_i} e^{-\beta W}}{\left(\prod_{i=1}^{s-1} Z_i\right)^{-1}}$$

$$= \sum_{E_1, E_2...E_{s-1}} \int dW \rho(W; E_1, E_2...E_{s-1}) e^{-\beta W} = \langle e^{-\beta W} \rangle, \tag{7a}$$

where $W$ is given by Eq. (3c), the sums run over all possible values of $E_1, E_2...E_{s-1}$, and

$$\rho(W; E_1, E_2...E_{s-1}) = \left(\prod_{i=1}^{s-1} Z_i\right)^{-1}$$
$$\times \int_{V_1, V_2...V_{s-1}} \prod_{i=1}^{s-1} dx_i \prod_{i=1}^{s-1} |\psi_{E_i}(x_i)|^2 \exp(-\beta \sum_{i=1}^{s-1} E_i) \exp(-\beta W) \tag{7b}$$
$$\times \delta\left(W - \sum_{i=1}^{s-1} [U(x_i, \lambda_{i+1}) - U(x_i, \lambda_i)]\right),$$

is a quantum work distribution function along an energy pathway characterized by $(E_1, E_2...E_{s-1})$. Thus, we can rewrite Eq. (6) as

$$\exp[-\beta \Delta F(\lambda_1, \lambda_s)] = \int dW \rho(W) \exp(-\beta W) = \langle \exp(-\beta W) \rangle, \tag{8}$$

where $\rho(W) = \sum_{E_1, E_2...E_{s-1}} \rho(W; E_1, E_2...E_{s-1})$ is the total work distribution function. In Eq. (8), $\rho(W)$ only exists for $s > 1$, and $\int dW \rho(W)$ is equal to unity [15].

Equations (6-8) suggest that if $s - 1$ complete sets of states $\{|E_i\rangle\}$ and eigenvalues $E_i$ of the coupled Hamiltonian are known, work distribution functions can be constructed to compute $\Delta F(\lambda_1, \lambda_s)$. We can also express the total work distribution function as

$$\rho(W) = \prod_{i=1}^{s-1} \int_{V_i} dx_i f(x_i) \delta\left(W - \sum_{i=1}^{s-1} [U(x_i, \lambda_{i+1}) - U(x_i, \lambda_i)]\right), \tag{9}$$

where $f_i(x_i) = \sum_{E_i} |\psi_{E_i}(x_i)|^2 \exp(-\beta E_i)/Z_i$ are the distribution functions of $x_i$. In computational studies, $f_i(x_i)$ is generated by sampling the quantum-thermal fluctuations of $x_i$, or by direct computation of the eigenstates and eigenvalues. For long enough relaxation times, $f_i(x_i)$ can be approximated as $\exp[-\beta k(x_i - \langle x_i \rangle)^2/2]$ to obtain

$$\Delta F_{\text{app}} = F(\lambda_s) - F_{\text{ref}}(\lambda_1, \lambda_s) = k\Delta \lambda \sum_{i=1}^{s-1} (\lambda_i - \langle x_i \rangle), \tag{10}$$

where $\langle x_i \rangle$ is the average value of $x_i$ at the $i$th pulling step [10]. For a sufficiently large number of pulling steps, the right-hand side of Eq. (10) becomes the Thermodynamic Integral [16], $\int_{\lambda_1}^{\lambda_s} \langle \partial \hat{H} / \partial \lambda \rangle_\lambda d\lambda$, where $\langle ... \rangle_\lambda$ is the average at each value of $\lambda$.

The expressions in Eqs. (7) imply that the work distribution function $\rho(W)$ is a sum of all possible quantum work distribution functions $\rho(W; E_1, E_2...E_{s-1})$. Thus, the predefined classical pathway $\lambda_1 \to \lambda_2 \to \lambda_3 ... \to \lambda_s$ is a sum of all possible quantum pathways. A quantum pathway is defined as $\{E_1 \to E_2 \to E_3 ... \to E_{s-1}$ and $x_1 \to x_2 \to x_3 ... \to x_{s-1}\}$. Based on this idea, we aim to identify those energy pathways $E_1 \to E_2 \to E_3 ... \to E_{s-1}$ which have the largest contribution to the free-energy change, given a set of $(x_1, x_2...x_{s-1})$. We also wish to determine which spatial pathways $x_1 \to x_2 \to x_3 ... \to x_{s-1}$ have the largest contribution to the free-energy change, given a set of $(E_1, E_2...E_{s-1})$. The first answer is to propose a possible picture of phase transitions or chemical reactions in terms of an energy diagram. The second one is to provide insight into how chemical reactions occur in the spatial domain.

To answer these questions, we use the variational principle for functionals [17]. The variational principle allows one to find optimal trajectories, which maximize or minimize an integral. It is noted that the left-hand side of Eq. (7a) contains the multi-dimensional integrals over $s-1$ variables $x_i$, one-dimensional integral over $W$ and sums (~ integrals in the continuum limit) over $s-1$ variables $E_i$. Given $(E_1, E_2 \ldots E_{s-1})$, the function under the integrals over variables $x_i$ can be expressed as $G_1(E_i, |\psi(x_i, E_i)|^2, W, x_i)$, where $E_i$, $|\psi(x_i, E_i)|^2 \equiv |\psi_{E_i}(x_i)|^2$ and $W$ are considered as functionals. Using the variational principle, we obtain

$$\frac{\partial \ln |\psi(x_i, E_i)|^2}{\partial E_i} = \beta(1 + \frac{\partial W}{\partial E_i}), \tag{11a}$$

where $i$ runs from 1 to $s-1$. Similarly, given $(x_1, x_2 \ldots x_{s-1})$ we express the function under the sums over variables $E_i$ as $G_2(|\psi(x_i, E_i)|^2, W, E_i,)$, where $x_i$ and $W$ are considered as functionals. Using the variational principle, we obtain another set of equations

$$\frac{\partial \ln |\psi(x_i, E_i)|^2}{\partial x_i} = \beta \frac{\partial W}{\partial x_i}. \tag{11b}$$

The solutions to Eq. (11a) are the spatial trajectories of optimal transitions $x_1 \to x_2 \to x_3 \ldots \to x_{s-1}$ given $(E_1, E_2 \ldots E_{s-1})$. The solutions to Eq. (11b) correspond to the energy trajectories of optimal transitions $E_1 \to E_2 \to E_3 \ldots \to E_{s-1}$ given $(x_1, x_2 \ldots x_{s-1})$. The trajectories satisfying both sets of Eqs. (11a-11b) yield the optimal contributions to the free-energy change. Integrating the above equations for a transition from the $(i-1)$th- to $i$th-pulling steps, we arrive at

$$\frac{|\psi_{E_i}(x_i)|^2}{|\psi_{E_{i-1}}(x_i)|^2} = e^{\beta(E_i - E_{i-1} + W_i - W_{i-1})}, \tag{12a}$$

$$\frac{|\psi_{E_i}(x_i)|^2}{|\psi_{E_i}(x_{i-1})|^2} = e^{\beta(W_i - W_{i-1})}, \tag{12b}$$

where $W_i = \sum_{j=1}^{j=i}[U(x_j, \lambda_{j+1}) - U(x_j, \lambda_j)]$. For the optimal trajectories, we combine the two sets of equations to obtain

$$\frac{|\psi_{E_i}(x_{i-1})|^2}{|\psi_{E_{i-1}}(x_i)|^2} = e^{\beta(E_i - E_{i-1})}, \tag{13}$$

which resembles the *detailed balance* equations $P_{i-1 \to i} / P_{i-1 \leftarrow i} = \exp(-\beta E_{i-1}) / \exp(-\beta E_i)$. In the *detailed balance* equations, $P_{i-1 \to i}$ is a forward transition probability for the system at $E_{i-1}$ moving into $E_i$, and $P_{i-1 \leftarrow i}$ is a reverse transition probability for the system at $E_i$ returning to $E_{i-1}$. Given $m$ pairs of $(x_{i-1}, x_i)$ satisfying Eq. (13), the sums of $|\psi_{E_i}(x_{i-1})|^2$ over $m$ values of $x_{i-1}$ and of $|\psi_{E_{i-1}}(x_i)|^2$ over $m$ values of $x_i$ are proportional to $P_{i-1 \to i}$ and $P_{i-1 \leftarrow i}$, respectively. Figure (2) illustrates an optimal transition (denoted by the blue arrow) based on Eq. (13), which is analogous to the Franck-Condon principle [18–20]. An optimal transition from $E_{i-1}$ to $E_i > E_{i-1}$ occurs, if $|\psi_{E_i}(x_{i-1})|^2 > |\psi_{E_{i-1}}(x_i)|^2$, which indicates that the probability for the particle at $E_i$ is more preferable than at $E_{i-1}$. It happens in the region $\Delta x$, which denotes the overlap between $|\psi_{E_{i-1}}(x_{i-1})|^2$ and $|\psi_{E_i}(x_i)|^2$. For example, in [21] the overlap $\Delta x$ among the electronic orbitals of two reactants is assumed in an oxidization-reaction theory, which involves electron transfers in solution. If $\Delta x$ is equal to zero, transitions not satisfying Eq. (13) are unlikely to occur. Consequently, the trajectories not satisfying Eq. (13) contribute much less than the optimal ones. Therefore, the optimal trajectories have the maximum contributions to the free-energy change.

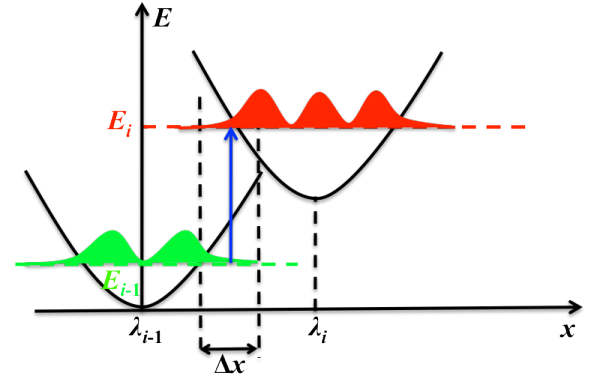

**Fig. 2.** Schematic diagram for optimal transitions. The parabolic curves represent an applied harmonic potential at the $(i-1)$th and $i$th pulling steps. The red-shaded and green-shaded areas are the probabilities $|\psi_{E_{i-1}}(x_{i-1})|^2$ and $|\psi_{E_i}(x_i)|^2$ respectively. The blue arrow indicates an optimal transition from $E_{i-1}$ (green-dashed line) to $E_i$ (red-dashed line). The region denoted by $\Delta x$ is the overlap between $|\psi_{E_{i-1}}(x_{i-1})|^2$ and $|\psi_{E_i}(x_i)|^2$.

To complete this discussion, we rewrite Eq. (8) in terms of dominant contributions to the free energy change. Using the principle of *detailed balance*, Jarzynski and Crooks [6,22] derived the JE to compute a free-energy change $\Delta F_S$ for stochastic processes. In our theory, $\Delta F_S$ exists as a term on the right-hand side of Eq. (8). For the optimal pathways following either Eq. (12a) or (12b) we consider them as deterministic because we can control work to induce reactions. We define the free-energy change for these pathways as $\Delta F_D$ and the free-energy change for the most optimal pathways as $\Delta F_{OP}$. Since the most optimal pathways have a contribution to both $\Delta F_S$ and $\Delta F_D$, we write the total free-energy change as

$$e^{-\beta \Delta F(\lambda_1, \lambda_s)} = e^{-\beta \Delta F_S} + e^{-\beta \Delta F_D} - e^{-\beta \Delta F_{OP}} + e^{-\beta \Delta F_B}, \tag{14}$$

where $\Delta F_B$ is due to biased pathways, which have a small contribution to the total free-energy change. If a sampling of reaction pathways does not capture any optimal pathways, which can be tested by Eqs. (12-13), then $\Delta F(\lambda_1, \lambda_s)$ is highly biased.

### III. TESTING

#### a. Control parameter $\lambda$ as the center of a harmonic potential

To test Eqs. (6-8), we compare $\Delta F(\lambda_1,\lambda)$ with the analytical free-energy changes by explicit calculation for an one-dimensional quantum harmonic oscillator. The non-perturbed Hamiltonian is given by $\hat{H}_0(\hat{p},\hat{x}) = \hat{p}^2/2m + k\hat{x}^2/2$, and a potential operator $\hat{U}(\hat{x},\lambda) = k(\hat{x}-\lambda)^2/2$ is applied. The eigenvalues of the coupled Hamiltonian $\hat{H}(\hat{p},\hat{x};\lambda) = \hat{H}_0(\hat{p},\hat{x}) + \hat{U}(\hat{x},\lambda)$ are $E_n = \hbar\omega(n+1/2) + k\lambda^2/4$, where $n$ is an integer and $\omega = (2k/m)^{1/2}$. The corresponding eigenstates are
$$\psi_n(x,\lambda) = \sqrt{(1/2^n n!)}\sqrt{m\omega/\pi\hbar}$$
$$\times \exp[-m\omega(x-\lambda/2)^2/2\hbar] H_n((x-\lambda/2)\sqrt{m\omega/\hbar}),$$
where the Hermite polynomials are $H_n(y) = (-1)^n \exp(y^2)\partial^n \exp(-y^2)/\partial y^n$. The analytical free-energy at each value of $\lambda$ is $F(\lambda) = (\hbar\omega/2)a^{-1}\ln(e^a - e^{-a}) + k\lambda^2/4$, where the reduced temperature is given by $a = \hbar\omega/2k_B T$.

We consider pulling protocols in which the values of $\lambda$ are changed from $\lambda_1 = 0$ to $\lambda_s = 1 \times (\hbar/m\omega)^{1/2}$ in increments of $\Delta\lambda = \lambda_s/(s-1)$, where $s$ is the number of pulling steps. The free-energy changes $\Delta F(\lambda_1,\lambda)$ computed from Eq. (8) are tested for $s \in \{2\text{-}11, 21\}$, $n$ numbers of eigenstates, eigenvalues $\in \{0\text{-}10, 20, 50\}$, and $a = 2^l$, with $l$ being integers, $-4 \leq l \leq 4$. The work distribution functions for each pulling step are evaluated by using the recursion relation (see Appendix B)
$$\rho_i(W) = Q_i \int dw \rho_{i-1}(w) f_{i-1}(\lambda_{i-1} + \frac{\Delta\lambda}{2} - \frac{W-w}{k\Delta\lambda}), \quad (15)$$
where $\rho_i(W)$ is the normalized work distribution at the $i$th-pulling step ($i > 1$), $f_i(x_i) = \sum_{j=0}^{n}|\psi_j(x_i,\lambda_i)|^2 \exp[-\beta E_j(\lambda_i)]$, $Q_i$ is the normalization factor of $\rho_i(W)$, and $\rho_2(W) = Q_2 f_1[\lambda_1 + \Delta\lambda/2 - W/k\Delta\lambda]$.

To illustrate these work distribution functions in quantum mechanics, we show them at $a = 1$, $n = 0$, and $s = 11$ in Fig. 3(a). From the work distribution functions, we compute the free-energy profile of $\Delta F(\lambda_1,\lambda)$ [shown in Fig. 3(b)]. $\Delta F(\lambda_1,\lambda)$ at $a = 1$, $n = 0$, and $s = 11$ perfectly fits with the exact free-energy profile, $\Delta F_{\text{Target}} = F(\lambda) - F(0) = (\hbar\omega/2)(\lambda/\sqrt{\hbar/m\omega})^2/4$. One can verify the perfect fit by analytically carrying out the integrations in Eq. (7a) (see Appendix C). By increasing $n$, $\Delta F(\lambda_1,\lambda_{11})$ converges to 0.241136 at $n = 7$ [see Fig. 3(c)]. Note that $F_{\text{ref}}(\lambda_1,\lambda_s)$ in Eq. (5) does not have the same expression as $F(\lambda_1)$ because of the non-zero commutation between the initial Hamiltonian and the applied potential operator. To examine the effect of the non-zero commutation, we show the dependence of $\Delta F(\lambda_1,\lambda_s)$ on $\Delta\lambda$ in Fig. 3(d). For sufficiently large $\Delta\lambda$ ($0.5\lambda_s$ to $1.0\lambda_s$), the non-zero commutation in Eq. (5) becomes significant. As $\Delta\lambda$ decreases, the commutation becomes negligible; hence $\Delta F(\lambda_1,\lambda_s)$ approaches $\Delta F_{\text{Target}}$ as a linear function of $\Delta\lambda$, $-0.0884 \Delta\lambda/\sqrt{\hbar/m\omega} + 0.2501$. This consistency suggests that the definition of $F_{\text{ref}}(\lambda_1,\lambda_s)$ [Eq. (5)] for the step-wise pulling protocol is valid at reduced temperature $a = 1$.

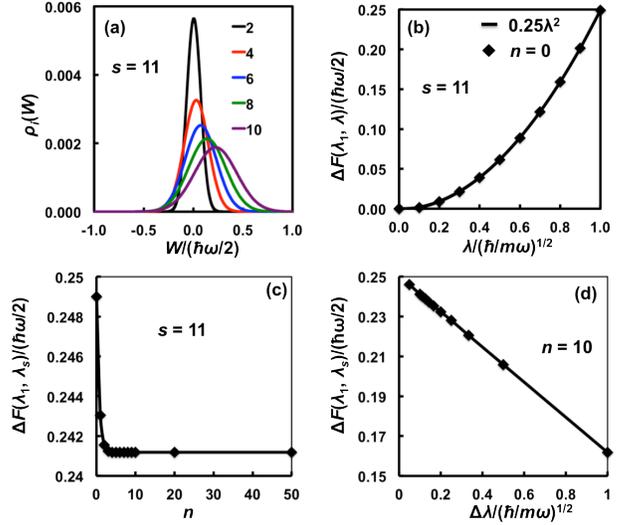

**Fig. 3.** (a) Normalized work distributions. The numbers indicate the $i$th-pulling steps at $s = 11$, and $n = 0$. (b) Free energy profiles at $s = 11$, and $n = 0$. (c) Free-energy changes $\Delta F(\lambda_1,\lambda_s)$ versus number $n$ of the eigenstates and eigenvalues. (d) Free-energy changes $\Delta F(\lambda_1,\lambda_s)$ versus increment $\Delta\lambda = \lambda_s/(s-1)$ at $n = 10$. All data here are evaluated at $a = 1$.

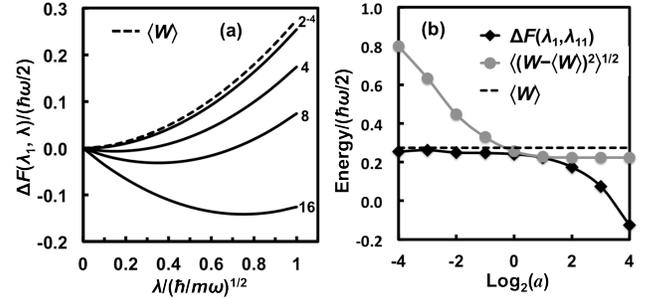

**Fig. 4.** (a) Free-energy profiles (solid lines) at $s = 11$, and for different reduced temperatures $a = \hbar\omega/2k_B T$. The numbers indicate the values of $a$. The dashed line is the average work $\langle W \rangle$ along the pathway. (b) Free-energy $\Delta F(\lambda_1,\lambda_{s=11})$, standard deviation $\langle (W-\langle W \rangle)^2 \rangle^{1/2}$ of work distributions, and average work $\langle W \rangle \approx 0.274 \hbar\omega/2$ versus $\log_2(a)$.

Fig. 4(a) shows the free-energy profiles (solid lines) at different temperatures for $\Delta\lambda = 0.1\lambda_{s=11}$, which are bounded above by the average mechanical work $\langle W \rangle$ (dashed line). We observe that $\langle W \rangle$ at $\lambda_{11}$ is unchanged ($\cong 0.274\hbar\omega/2$) over the wide range of temperatures studies here [see Fig. 4(b)]. The standard deviation $\langle (W-\langle W \rangle)^2 \rangle^{1/2}$ of the work distributions decreases with temperature, and converges to $0.2236\hbar\omega/2$. The unchanged value of $\langle W \rangle$ and the convergence of $\langle (W-\langle W \rangle)^2 \rangle^{1/2}$ at low temperatures indicate that the work distribution is practically independent of temperature (at $s = 11$). This indicates that the ground state ($n = 0$) at each pulling step significantly contributes to the work distribution at sufficiently low temperatures.

At high temperatures ($a < 1$), the free-energy profiles are not noticeably distinguishable from those at $a = 1$, but more energy levels and wave functions contribute to the work distributions. As shown in Fig. 4(b), $\Delta F(\lambda_1,\lambda_{11})$ at $\lambda_{11} =$ $1 \times (\hbar/m\omega)^{1/2}$ converges to $0.25 \pm 0.01$ ($\hbar\omega/2$). The error ($\approx 0.01/0.25 = 4\%$) is due to truncated number $n$ of wave functions. However, $\Delta F(\lambda_1,\lambda_{11})$ becomes negative when lowering the temperature ($a > 1$). The negative values of $\Delta F(\lambda_1,\lambda_{11})$ indicate the strong effect of the non-zero commutation between the initial Hamiltonian and the applied potential at a finite number $s = 11$ of pulling steps. This means that not only $F(\lambda_i)$, but also $F_{\text{ref}}(\lambda_1,\lambda_i)$ varies as $i$ runs from 2 to $s = 11$. Since only the ground state at each pulling step significantly contributes to the work distribution function at low temperatures, the free-energy changes can be estimated from $\Delta F(\lambda_1,\lambda_s) = k\Delta\lambda^2(s-1)^2[1-(a-1)/(s-1)]/4$ (see Appendix C). As $a > s$, the estimated $\Delta F(\lambda_1,\lambda_s)$ becomes negative, as observed in Figs. 4(a-b). If $s$ is much larger than $a$, the estimated $\Delta F(\lambda_1,\lambda_s)$ approaches $\Delta F_{\text{Target}}$. As a result, at sufficiently low temperatures (large $a$) and very small increments $\Delta\lambda$ the variation of $F_{\text{ref}}(\lambda_1,\lambda_i)$ with respect to $\lambda_i$ is significant at steps $i \sim a$, but becomes negligible at steps $i \gg a$. Therefore, we can identify the negative values of $\Delta F(\lambda_1,\lambda_i)$ at steps $i \sim a$ as a quantum effect of the JE at low temperatures.

### b. Control parameter $\lambda$ as the spring constant of a harmonic applied potential

To directly compare work distribution functions and free-energy profiles computed by Eqs. (6-8) with those derived by using the JE for closed quantum systems [5], we now vary the spring constant following a step-wise protocol. In this case, $\hat{H}_0(\hat{p},\hat{x}) = \hat{p}^2/2m + (k_0 - \Delta k)\hat{x}^2/2$, $\hat{U}(\hat{x},k_i) = k_i\hat{x}^2/2$, where $k_i$ is $i\Delta k$, $\Delta k$ is an increment, and $s$ is the number of pulling steps. The eigenvalues of the coupled Hamiltonian $\hat{H}(\hat{p},\hat{x};k_i) = \hat{H}_0(\hat{p},\hat{x}) + \hat{U}(\hat{x},k_i)$ are $E_n(\omega_i) = \hbar\omega_i(n+1/2)$, where $n$ is an integer, $\omega_i$ is given by $\omega_0[1+(i-1)\delta]^{1/2}$, $\omega_0 = (k_0/m)^{1/2}$, and $\delta = (\omega_s^2 - \omega_0^2)/\omega_0^2(s-1) = \Delta k/k_0$. We choose $\omega_s = 1.3\omega_0$ as used in Ref. [5]. The corresponding eigenstates are $\psi_n(x,\omega_i) = \sqrt{(1/2^n n!)}\sqrt{m\omega_i/\pi\hbar}\exp[-m\omega_i x^2/2\hbar]$ $\times H_n(x\sqrt{m\omega_i/\hbar})$, using the Hermite polynomials $H_n(y) = (-1)^n\exp(y^2)\partial^n\exp(-y^2)/\partial y^n$. The analytical free-energy at each value of $\omega_i$ is $F(\omega_i) = \sinh(a_0[1+(i-1)\delta]^{1/2}/2)$, where the reduced temperature is again given by $a_0 = \hbar\omega_0/k_B T$.

First, we examine a case at high temperature, e.g., $a_0 = 0.1$ (used in Ref. [5]). The work distribution functions $\rho_i(W)$ are computed from $f_i(x_i) = \sum_{j=0}^n |\psi_j(x_i,\omega_i)|^2 \exp[-\beta E_j(\omega_i)]$, where $n = 100$ is chosen. The expression of work $W$ is computed from $\sum_{i=1}^{s-1} 0.5\delta/\sqrt{1+\delta(i-1)}(x_i^2 m\omega_i/\hbar)$, which is always non-negative. Fig. 5(a) shows the work distribution function $\rho_2(W)$ for free-energy difference $\Delta F(\omega_1=\omega_0,\omega_s)$ at $s = 2$. $\rho_2(W)$ has a sharp peak at $W = 0$, which resembles the feature for adiabatic processes, $\sim \exp[-\beta W\omega_0/(\omega_s-\omega_0)]\theta(W)$, where $\theta(W)$ is the Heaviside step function of $W$. By increasing $s$ to 11, this peak in the work distribution $\rho_{11}(W)$ for the same $\Delta F(\omega_1=\omega_0,\omega_s)$ smoothens out [see Fig. 5(b)]. The free-energy profiles shown in Fig. 5(c) for $s = 2$ and 11 agree with the targeted $\Delta F_{\text{Target}} = -\beta^{-1}\ln[F(\omega_1=\omega_0)/F(\omega_s)]$. Note that the work distributions do not have negative tails which were observed in Ref. [5] for non-adiabatic processes. By definition, the values of work are only negative for the cases of reducing the spring constant ($\delta < 0$), which can result into lowering the temperature of the system.

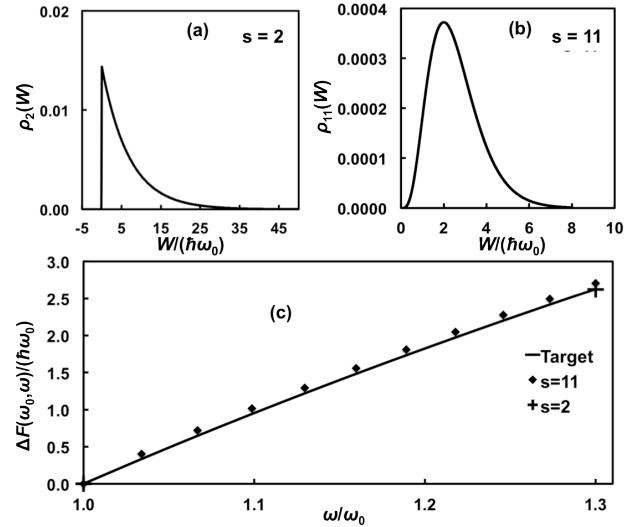

**Fig. 5.** (a) Normalized work distribution at $s = 2$. (b) Normalized work distribution function at $s = 11$. (c) Free-energy profiles. The work distributions and free-energy profiles are computed at $a_0 = \hbar\omega_0/k_B T = 0.1$.

Last but not least, we compute free-energy changes at very low temperatures. Analogous to the previous example, we observe that at low temperatures only the ground state at each pulling step significantly contributes to the work distribution functions. This observation allows us to analytically estimate $\Delta F(\omega_1,\omega_s)$. Similar to the procedure in Appendix C, we arrive at

$$\Delta F(\omega_1,\omega_s) = \frac{\hbar\omega_0}{2a_0}\sum_{i=1}^{s-1}\ln(1+\frac{a_0\delta}{2\sqrt{1+\delta(i-1)}})$$
$$\cong \sum_{i=1}^{s-1}\frac{\hbar\omega_0\delta}{4\sqrt{1+\delta(i-1)}}+O((a_0\delta)^2)$$
$$\cong \hbar\omega_0\int_0^{(\omega_s^2-\omega_0^2)/\omega_0^2}\frac{dy}{4\sqrt{1+y}}+O(1/s)+O((a_0\delta)^2) \quad (16)$$
$$= \frac{\hbar(\omega_s-\omega_0)}{2}+O(1/s),$$

which is consistent with the low-temperature limit in Ref. [5].

## IV. DISCUSSIONS AND CONCLUSIONS

It is noted that the step-wise pulling protocol and Eqs. (6-7) have some similarities with the decomposition scheme applied in the Trotter-Suzuki formula [23,24], $e^{-\beta(\hat{H}+\hat{U})} = \lim_{q\to\infty}(e^{-\beta\hat{H}/q}e^{-\beta\hat{U}/q})^q$, where $\hat{H}$ and $\hat{U}$ are the initial Hamiltonian and applied potential operators, and $\hat{U}$ is fixed. The decomposition is useful to derive the path integral representation of the partition function [25,26]. The partition function can be a multi-dimensional integral over $q$ sets of a coordinate variable $x$, i.e., $x_j$ with $j$ being 1 to $q$. Similarly, Eq. (7a) contains a multi-dimensional integral over $s-1$ sets of a reaction-coordinate variable $x_i$. In Eqs. (6-7), the operators $\delta\hat{W}_i = \hat{U}(\hat{x}, \lambda_{i+1}) - \hat{U}(\hat{x}, \lambda_i)$ and the mechanical work $W$ weigh those rare reaction pathways, which have the largest contribution to free-energy changes. A reaction pathway is defined as two series of energy pathways ($E_1, E_2...E_{s-1}$) and spatial pathways ($x_1, x_2...x_{s-1}$). If $s$ is small, the accuracy of the pathways is poor because the energy and spatial differences between successive pulling steps are large. Equations (12-13) (see Fig. 2) suggest a criterion that certain overlaps of successive wave functions are sufficient to reconstruct reliable rare reaction pathways. Thus, $s$ can be finite at a certain range of temperatures (e.g. larger than Debye temperatures). But at low temperatures (see Sec. III), an infinitely large number of $s$ (like for $q$) will be needed to guarantee the convergence of free-energy changes.

This theory of the JE for quantum systems is a generalization from the proof for classical systems. As a result, the same discretized mechanical work can be applied to any classical and quantum systems. Unlike the JE for classical systems, a work distribution function generated from Eqs. (7, 9) gives the difference between a final free energy $F(\lambda_s)$ and the reference free energy $F_{\text{ref}}(\lambda_1, \lambda_s)$. In the classical limit, $F_{\text{ref}}(\lambda_1, \lambda_s)$ becomes $F(\lambda_1)$. For a one-dimensional harmonic oscillator at low temperature, the variation of $F_{\text{ref}}(\lambda_1, \lambda_s)$ due to quantum effects becomes significant at small $s$, but negligible at very large $s$ (see Sec. III). Thus, the definition of $F_{\text{ref}}(\lambda_1, \lambda_s)$ still ensures the convergence of $\Delta F(\lambda_1, \lambda_s)$ by collecting many fluctuations of a reaction coordinate along a given pathway. The fluctuations are affected by the rest of systems and heat baths, which can be arbitrarily coupled in any dynamics. They can be simply computed from wave functions weighted by exponentials of corresponding eigenvalues, and then used for constructing work distribution functions. In the presence of heat baths, the fluctuations can be approximated as $\exp[-\beta k(x_i - \langle x_i \rangle)^2/2]$ to arrive at Eq. (10), which is consistent with the Thermodynamic Integral [16], $\int_{\lambda_1}^{\lambda_s} \langle \partial \hat{H}/\partial \lambda \rangle_\lambda d\lambda$. Therefore, the convergence and consistency suggest the validity of the quantum expressions of the JE, Eqs. (6-8).

The introduction of the discretized mechanical work $W$ into Eqs. (6-8) is useful to determine which reaction pathways are optimal. Without $W$, it is unclear to prove that the most optimal reaction pathways follow the principle of *detailed balance*, $P_{i-1 \rightarrow i} / P_{i-1 \leftarrow i} = \exp(-\beta E_{i-1})/\exp(-\beta E_i)$. We also pointed out that the transition probabilities $P_{i-1 \rightarrow i}$ and $P_{i-1 \leftarrow i}$ should contain information of wave functions along any optimal transition pathway, although they can be arbitrarily defined in practice [26]. In the paper by Metropolis *et al.* [27], the idea of using the transition probabilities satisfying the principle of *detailed balance* is to quickly drive systems into equilibrium states for classical systems; hence sampling of canonical distributions can be done less costly. The sampling based on the transition probabilities might be enhanced or smoothed by using Eqs. (12-13), which restrict the overlaps between successive states. Addressing the advantages of using Eqs. (12-13) and time evolution of the transitions would be of interest for future research. Since the time evolution has not yet been discussed in the theory and test case, it is also essential to estimate the relaxation time to characterize the limits of non-equilibrium processes, as examined in classical systems [10].

On one hand, from the JE we have obtained *detailed balance* for the most optimal transition pathways. On the other hand, using the principle of *detailed balance* Jarzynski and Crooks [6,22] derived the JE for classical stochastic processes. Furthermore, Boltzmann proved that the principle of *detailed balance* sets a sufficient condition for entropy growth [28]. As a result, we infer that the entropy growth follows the most optimal transition pathways for any dynamics.

One possible consequence of this theory is to suggest an approach to test the JE in Bose-Einstein condensates (BECs) [9]. The major difficulty of testing the JE is to measure work distribution functions in microscopic systems without interfering with the quantum dynamics. Since an applied harmonic potential can be used to trap BECs [29], it is possible to take advantage of that potential to perform work to observe how the condensate particles interact with the rest. The conventional way of constructing work distribution functions is to perform as many pulling trajectories as possible at a certain pulling speed. In other words, one might have to replicate many BECs and examine the effects of the pulling speed like in the case of unfolding RNAs [30]. One challenge of the conventional way is to know how many pulling trajectories are sufficient. If the potential is monitored using a step-wise pulling protocol, in which at each pulling step the system is relaxed long enough so that the quantum dynamics is not destroyed in a single step-wise pulling trajectory, then the work distribution functions can be constructed from the distributions of the fluctuating center-of-mass of the condensate particles. Burger *et al.* [31] showed in an experiment that for small increments (displacements) $\Delta\lambda \le 50$ $\mu m$, BECs have oscillating frequencies shifted and are undamped on time scales of milliseconds. Motivated by the experiment, we suggest that work distribution functions for Eq. (1) can be constructed from the distributions of the center-of-mass of the oscillating BECs deduced from absorption images [31]. Compared to other proposed methods, e.g., trapped ions in a linear Paul trap [32] or heat-transfer measurements [33], our approach provides an alternative perspective and insight into the equilibration dynamics of quantum systems.

In summary, we have proposed a generalization of Jarzynski's Equality [Eqs. (6-8)] for quantum systems based on a step-wise pulling protocol. We showed that the mechanical work in Eq. 3(c) can be used to generate work distribution functions, and evaluate free-energy changes via Eqs. (6-8). The work distribution functions can be constructed from (1) eigenstates and eigenvalues of a coupled Hamiltonian operator, or (2) by collecting the thermal and quantum fluctuation distributions of a reaction coordinate.

Using a simple harmonic potential to perform work and based on the variational principle, we derived two sets of equations to identify optimal transition pathways. The optimal transition pathways satisfying both sets of the equations were found to follow the principle of *detailed balance*. Finally, we tested the theory by explicit analysis of a quantum harmonic oscillator, computing free-energy changes using Eqs. (6-8). At temperatures $T \sim \hbar\omega/2k_B$, the convergence of the free-energy changes requires a finite number of many eigenstates and eigenvalues as small as 7, for a step-wise increment along the reaction pathway as small as $0.1 \times (\hbar/m\omega)^{1/2}$. By varying the angular frequency, we obtained the same limits derived from the JE for closed systems. At low temperatures, the ground state at each pulling step dominantly contributes to the work distribution function, and a large number $s$ of pulling steps is required to have convergent free-energy profiles.

We would like to thank Lorenzo Campos Venuti and Paolo Zanardi for useful conversations, and acknowledge financial support by the Department of Energy grant DE-FG02-05ER46240. We also acknowledge Sebastian Deffner and Eric Lutz for their suggestions.

## APPENDIX A

Here, we consider the relation between $\hat{O}_{\text{total}}$ and $\hat{W}$ in Eqs. (3). Let's examine the relation in classical systems [10], where the non-operator Hamiltonian is similar to Eq. (2). Then, the expression of $O_{\text{total}}$ (expectation value) is computed as

$$O_{\text{total}} = \int_{t_0}^{t_s} \frac{\partial H(p^{3N}, r^{3N-1}, x; \lambda_1, \lambda_2 \ldots \lambda_s)}{\partial t} dt$$
$$= U(x_0, \lambda_1) - U(x_s, \lambda_s) + \sum_{i=1}^{s-1} [U(x_i, \lambda_{i+1}) - U(x_i, \lambda_i)] \quad \text{(A.1)}$$
$$= U(x_0, \lambda_1) - U(x_s, \lambda_s) + W,$$

where

$$U(x_i, \lambda_{i+1}) = \int U(x, \lambda_{i+1})[\partial \theta(t - t_i)/\partial t] dt,$$
$$U(x_i, \lambda_i) = -\int U(x, \lambda_i)[\partial \theta(t_i - t)/\partial t] dt, \quad \text{(A.2)}$$

$W$ is given by Eq. (3c), and $x_i$ is a value of the reaction coordinate $x$ at time $t_i$. The relation between $O_{\text{total}}$ and $W$ is as clear as in Eq. (A.1). A value of $O_{\text{total}}$ indicates absorption energy along a pathway $x_0 \to x_1 \ldots \to x_s$ and $W$ is mechanical work performed by the applied potential. Since time $t_i$ arbitrarily begins from $t_{i-1}$, a thermal fluctuation distribution of $x_i$ during relaxation time $t_i - t_{i-1}$ exists and plays the same role as the single value of $x_i$ at time $t_i$. Thus, the idea of using all possible values of $x_i$ during relaxation time (instead of one value at time $t_i$) is consistent with the ergodic hypothesis of thermodynamics, which implies the equivalence between the averages over time ($t_i$) and over phase space ($x_i$) represented by Hamiltonian $H_i(p^{3N}, r^{3N-1}, x; \lambda_i)$. Based on this idea, one can construct work distribution functions from the thermal fluctuation distributions of $x_i$ [10].

However, in quantum mechanics one cannot pass time $t_i$ to operator $\hat{x}$ as to expectation value $x$ in Eqs. (A.2). As a result, operator $\hat{O}_{\text{total}}$ is identical to zero and does not have any physical meaning. Strictly speaking, mechanical work operator $\hat{W}$ does not indicate useful physical meanings either, but its expectation values defined by Eq. (3c). Thus, the relation should be understood in terms of expectation values of $\hat{O}_{\text{total}}$ and $\hat{W}$ over spatial domains along a pathway.

## APPENDIX B

To derive Eq. (15), we start from Eq. (9). First, let $W_i$ be $\sum_{j=1}^{j=i}[U(x_j, \lambda_{j+1}) - U(x_j, \lambda_j)]$ for $1 \leq i \leq s - 1$, and $\rho_{i+1}(W)$ be the work distribution of $W_i$. Then we have a simple relation between $W_i$ and $W_{i-1}$:

$$W_i = W_{i-1} + \frac{k\Delta\lambda}{2}(2\lambda_i + \Delta\lambda - 2x_i), \quad \text{(B.1)}$$

where $\Delta\lambda = \lambda_s/(s-1)$, $W_0 = 0$, and $\rho_1(W) \equiv 1$ [15]. As a result of Eq. (9) for $i = 1$, the un-normalized work distribution function $\rho_2(W)$ is equal to $f_1[\lambda_1 + \Delta\lambda/2 - W/k\Delta\lambda]$. For $i = 2$, the un-normalized work distribution function $\rho_3(W)$ is

$$\rho_3(W) = \int_{-\infty}^{\infty} dx_1 \int_{-\infty}^{\infty} dx_2 f_1(x_1) f_2(x_2)$$
$$\times \delta[W - W_1 - \frac{k\Delta\lambda}{2}(2\lambda_2 + \Delta\lambda - 2x_2)]. \quad \text{(B.2)}$$

By changing the variable $x_1 = \lambda_1 + \Delta\lambda/2 - W_1/k\Delta\lambda$, Eq. (B.2) becomes

$$\rho_3(W) = \int_{-\infty}^{\infty} dW_1 \int_{-\infty}^{\infty} d(x_2/k\Delta\lambda) \rho_2(W_1) f_2(x_2)$$
$$\times \delta[W - W_1 - \frac{k\Delta\lambda}{2}(2\lambda_2 + \Delta\lambda - 2x_2)]. \quad \text{(B.3)}$$

Integrating Eq. (B.3) over $x_2$, and changing the variable $W_1$ to $w$, we obtain

$$\rho_3(W) = (1/k\Delta\lambda) \int_{-\infty}^{\infty} dw \rho_2(w) f_2(\lambda_2 + \frac{\Delta\lambda}{2} - \frac{W-w}{k\Delta\lambda}). \quad \text{(B.4)}$$

For $i > 2$, one can easily verify

$$\rho_i(W) = \int_{-\infty}^{\infty} dx_{i-1} \times$$
$$\int_{-\infty}^{\infty} dw \{[\int_{-\infty}^{\infty} dx_1 \ldots \int_{-\infty}^{\infty} dx_{i-2} f_1(x_1) \ldots f_{i-2}(x_{i-2}) \delta(w - W_{i-1})]$$
$$\times f_{i-1}(x_{i-1}) \delta[W - w - \frac{k\Delta\lambda}{2}(2\lambda_{i-1} + \Delta\lambda - 2x_{i-1})]\}$$
$$= \int_{-\infty}^{\infty} dx_{i-1} \int_{-\infty}^{\infty} dw \rho_{i-1}(w) f_{i-1}(x_{i-1})$$
$$\times \delta[W - w - \frac{k\Delta\lambda}{2}(2\lambda_{i-1} + \Delta\lambda - 2x_{i-1})] \quad \text{(B.5)}$$
$$= (1/k\Delta\lambda) \int_{-\infty}^{\infty} dw \rho_{i-1}(w) f_{i-1}(\lambda_{i-1} + \frac{\Delta\lambda}{2} - \frac{W-w}{k\Delta\lambda}),$$

which is Eq. (15), and the normalization factor $Q_i$ is computed as $Q_i = 1/\int_{-\infty}^{\infty} dW \rho_i(W)$.

## APPENDIX C

Here, we analytically carry out the integrations in Eq. (7a) for one-dimensional harmonic oscillator with $n = 0$, which can be written as

$$e^{-\beta \Delta F(\lambda_1, \lambda_s)} = \frac{\prod_{i=1}^{s-1} \int dx_i |\psi_0(x_i, \lambda_i)|^2 \exp[-\beta(\frac{\hbar\omega}{2} + \frac{k\lambda_i^2}{4})]\exp[-\beta \delta W_i]}{\prod_{i=1}^{s-1} \int dx_i |\psi_0(x_i, \lambda_i)|^2 \exp[-\beta(\frac{\hbar\omega}{2} + \frac{k\lambda_i^2}{4})]}, \quad (C.1)$$

where $\delta W_i = k\frac{\Delta\lambda}{2}(2\lambda_i + \Delta\lambda - 2x_i)$, $\beta$ is $1/k_B T$, $\omega = [2k/m]^{1/2}$, $\psi_0(x, \lambda) = \sqrt[4]{m\omega/\pi\hbar} \exp[-m\omega(x - \lambda/2)^2/2\hbar]$, $\lambda_i$ is $(i-1)\Delta\lambda$, and $\Delta\lambda = \lambda_s/(s-1)$. Note that the factors $\exp[-\beta(\frac{\hbar\omega}{2} + \frac{k\lambda_i^2}{4})]$ cancel out in Eq. (C.1), and the integrations in the denominator are equal to unity. By converting $x_i$, $\lambda_i$, and $\Delta\lambda$ into dimensionless variables (in units of $\sqrt{\hbar/m\omega}$), we simplify Eq. (C.1) to

$$e^{-\beta \Delta F(\lambda_1, \lambda_s)} = \prod_{i=1}^{s-1} \int dx_i \exp[-(x_i - \frac{\lambda_i}{2})^2 - \frac{2x_i a\Delta\lambda}{2}]$$
$$\times \exp[-\frac{a\Delta\lambda}{2}(2\lambda_i + \Delta\lambda)]$$
$$= \prod_{i=1}^{s-1} \int dx_i \exp[-\left(x_i - \frac{\lambda_i + a\Delta\lambda}{2}\right)^2]$$
$$\times \exp[-\frac{a\Delta\lambda}{2}(2\lambda_i + \Delta\lambda) + \left(\frac{a\Delta\lambda}{2}\right)^2 + \frac{a\Delta\lambda\lambda_i}{2}]$$
$$= \exp\left[-\frac{a\Delta\lambda^2(s-2)(s-1)}{4} + (s-1)\left(\frac{a^2\Delta\lambda^2}{4} - \frac{a\Delta\lambda^2}{2}\right)\right]$$
$$= \exp\left[-\frac{a\Delta\lambda^2(s-1)(s-a)}{4}\right], \quad (C.2)$$

where $a = \hbar\omega/2k_B T$. As $a = 1$, we obtain $\Delta F(\lambda_1, \lambda_s) = k(s-1)^2\Delta\lambda^2/4$, which is $k\lambda_s^2/4 = \Delta F_{Target}$.